\documentclass[preprint,authoryear,12pt]{elsarticle}

\biboptions{authoryear}

\usepackage{graphicx}
\usepackage{float}
\usepackage{subfig}
\usepackage{enumitem}

\usepackage{amssymb}

\usepackage[ps2pdf,%
a4paper=true,%
breaklinks=true,%
colorlinks=true,%
pdfauthor={First Author et al.},%
pdftitle={Template for manuscripts in Advances in Space Research}%
]{hyperref}

\journal{Advances in Space Research}

\begin{document}

%%%%%%%%%%%%%%%%%%%%%%%%%%%%%%%%%%%%%%%%%%%%%%%%%%%%%%%%%%%%%%%%%%%%%%%%%%%%%
%% Frontmatter
\begin{frontmatter}


% Use the tnoteref command within \title and fnref within \author or \address for footnotes;
% use the corref command within \author for corresponding author footnotes;
% \author{Name\corref{cor1}\fnref{label2}}
% \address{Address\fnref{label3}}

\title{RadioAstron probes the ultra-fine spatial structure in the H$_2$O maser emission in the star forming region~W49N}
%\tnotetext[footnote1]{This template can be used for all publications in Advances in Space Research.}

% Use optional labels to link authors explicitly to addresses:

\author[asc,urfu]{N.N. Shakhvorostova\corref{cor}}
\cortext[cor]{Corresponding author}
%\fntext[footnote2]{Additional information regarding the corresponding author}
\ead{nadya@asc.rssi.ru}

\author[urfu]{A.M. Sobolev}

\author[camb]{J.M.~Moran}

\author[asc]{A.V. Alakoz}

\author[kag]{H.~Imai}

\author[asc]{V.Y.~Avdeev}

\address[asc]{Astro Space Center, Lebedev Physical Institute, Russian Academy of Sciences, 84/32 Profsoyuznaya st., Moscow, GSP-7, 117997, Russia}
\address[urfu]{Astronomical Observatory, Institute for Natural Sciences and Mathematics, Ural Federal University, 19 Mira street, Ekaterinburg, 620002, Russia}
\address[camb]{Harvard-Smithsonian Center for Astrophysics, 60 Garden Street, Cambridge, MA 02138, USA}
\address[kag]{Center for General Education, Kagoshima University, 1-21-30 Korimoto, Kagoshima 890-0065, Japan}

\begin{abstract}

H$_2$O maser emission associated with the massive star formation region W49N were observed with the Space-VLBI mission \textit{RadioAstron}. The procedure for processing of the maser spectral line data obtained in the \textit{RadioAstron} observations is described. Ultra-fine spatial structures in the maser emission were detected on space-ground baselines of up to 9.6 Earth diameters. The correlated flux densities of these features range from 0.1\% to 0.6\% of the total flux density. These low values of correlated flux density are probably due to turbulence either in the maser itself or in the interstellar medium.
\end{abstract}


\end{frontmatter}

\parindent=0.5 cm

%%%%%%%%%%%%%%%%%%%%%%%%%%%%%%%%%%%%%%%%%%%%%%%%%%%%%%%%%%%%%%%%%%%%%%%%%%%%%
%% Main text
\section{Introduction}
\label{intro}

Water vapor masers are common tracers of sites of star formation in their early phases, and therefore maser observations are a crucial tool in the study of the star formation process. Since H$_2$O masers exhibit compact structures in these regions, study of their angular structure require extremely high angular resolution. Typical scales on which H$_2$O masers occur are about 1-100 AU (for example, see \citealt{Imai2002}) assuming that they are unresolved in milli-arcsec angular resolution or finer. The highest angular resolution can be provided by making use of the Space-VLBI technique of \textit{RadioAstron} allowing investigation of the most compact structures in star forming regions comparable to the sizes of those H$_2$O masers.

\textit{RadioAstron} is an international space VLBI project involving the 10-m Space Radio Telescope (SRT) on board the satellite \textit{Spektr-R} in cooperation with many ground radio telescopes~(for more details see \citealt{Kardashev2013}). The SRT was launched in 2011 on an elliptical orbit whose plane is evolving with time with an apogee of up to 370~000 km. It operates at frequencies of 22, 5, 1.6, and 0.3~GHz~\footnote{ www.asc.rssi.ru/radioastron/index.html, last modified on January 2019.}. 

The high-mass star-forming region W49N (G43.16+00.01) is a part of W49A, which is the most massive and luminous star-forming complex in our Galaxy (see, for example, \citealt{Sievers1991}). W49N is located in the Perseus arm near the solar circle in the first Galactic quadrant at the distance of 11.11$\pm$0.8 kpc from the Sun \citep{Zhang2013}. The distance to W49N from the Galactic mid-plane is only 3~pc. This region contains numerous 22~GHz H$_2$O masers that form the most luminous maser set ($\sim1L_{\odot}$) in the Galaxy. This makes W49N an excellent target for \textit{RadioAstron} maser observations. W49N was observed as part of the \textit{RadioAstron} maser survey during 2014-2015~\citep{sobolev2017}. The highest angular resolution of 23~$\mu$as for galactic masers was achieved in these observations on a baseline of 9.6 Earth diameters (ED).

\section{Observations}
\label{obs}

We conducted three observing sessions on W49N as listed in the Table~\ref{obssummary}. Each observing segment was split into scans of 10-20 minutes. The SRT data were transmitted in real time to the ground tracking stations in Pushchino (Russia) or Green Bank (USA) depending on the SRT visibility conditions \citep{Kardashev2013,Ford2014}. Left- and right-hand circular polarization data were recorded with a total bandwidth of 32~MHz per polarization. The frequency coverage in the first and second experiments was 22.212$-$22.244~GHz and in the third experiment~-- 22.220$-$22.252~GHz. The spectral setup between second and third observations was changed in order to put a new flaring feature around $-63$~km/s in the central part of the upper side band (see Section~4). The $uv$-plane coverage for the second observing session is shown on Figure~1. The antenna pointing and phase-tracking center for all the scans on W49N in Table~\ref{obssummary} was set to RA$=19^h10^m13.4096^s$, DEC$=09^{\circ}06^{\prime}12.803^{\prime\prime}$, where the most luminous UCHII region~G in the W49N complex is located \citep{Dreher1984}. 

There was a calibration scan for 5~minutes provided for ground
telescopes prior to the beginning of each observing segment.
No calibrators were observed on the SRT due to technical and
other restrictions in these three experiments. According to \citep{Kovalev2019} and \textit{RadioAstron} User Handbook~\footnote{ The RadioAstron User Handbook, ver.~2.92, March~12,~2018,\\ http://www.asc.rssi.ru/radioastron/documents/rauh/en/rauh.pdf.} there are no bright calibrators at 1.3 cm in the close vicinity of W49N, compact enough to give interferometric fringes with RadioAstron on long baselines (up to almost 10 Earth diameters in case of our experiments). However, observations of such calibration sources were not necessary for the present study.

\begin{table}
\scriptsize
\begin{center}
\caption{Summary of observing sessions.}
\label{obssummary}
\begin{tabular}{cclcccc}
\hline \hline

N$^\circ$ & Epoch & Ground telescope  & Frequency and & Obs. & Baseline   & Fringe   \\
          &       &   array           & velocity range, & time,     & length,     & spacing,  \\
          &       &                   & GHz / km/s     & min  &  ED / km & $\mu$as \\    
\hline
1 & 18 May 2014   & Effelsberg 100-m & 22.212 $-$ 22.244 &  60  & 3.0 & 74  \\
  &               &                  & $-$81.80 $-$ +349.39  &          & $\sim$38 000 & \\
              \hline
2 & 27 April 2015 & Effelsberg 100-m & 22.212 $-$ 22.244 & 60 & 9.6 & 23\\
   &           & Yebes 40-m & $-$81.80 $-$ +349.39 & & $\sim$122 000 & \\
   &           & Torun 32-m & & & & \\
    &          & Hartebeesthoek 26-m & & & & \\
              \hline
3 & 22 May 2015   & Green Bank 100-m & 22.220 $-$ 22.252 & 50 & 8.6 &  26\\
     &         & Effelsberg 100-m & $-$190.65 $-$ +240.54 &    & $\sim$110 000 & \\
      &        & Medicina 32-m & &  & & \\
       &       & Toru\'{n} 32-m &  & & & \\
\hline
\hline
\end{tabular}
\end{center}
\end{table}

\section{Data reduction}
\label{reduction}

The data were correlated by the \textit{RadioAstron} correlator developed at the Astro Space Center (ASC) in Moscow \citep{Likhachev2017}. Spectral cross-correlation of the data were obtained with a frequency resolution of 7.81~kHz. The integration time of 0.125~seconds provided a wide field of view  of about 10 arcsec. 

Post-correlation data reduction was performed using the PIMA software~\footnote{ VLBI processing software PIMA, http://astrogeo.org/pima, last modified on March 8, 2019.} \citep{Petrov2011}. Its advanced fringe fitting algorithm was developed in order to find fringes on weak radio sources, which is particularly helpful in case of the very long baseline \textit{RadioAstron} observations.
PIMA performs fringe searches in a narrow part of the spectrum corresponding to certain maser features, and thus, for providing solutions of phases, group delays, fringe rates, and phase for the next iteration of data correlation. 

As was noted in Section~\ref{obs}, no calibrators were observed with the SRT. This made it impossible to determine the exact residual group delays for space baselines. Nevertheless, if the maser line gives bright fringes and if it is relatively wide, this yield the necessary signal to noise to permit an approximate delay determination.
Since many features in W49N are known to have apparent sizes of $\sim$250~$\mu$as \citep{Gwinn1988a}, it was expected that only the most compact components would remain on long space baselines. We used PIMA for delay solutions, and then the final data set was processed using a standard astronomical package AIPS~\footnote{ Astronomical Image Processing System, http://www.aips.nrao.edu, last modified on March 6, 2019.}.
%Since a single maser spectral feature is usually a blend of several features corresponding to different emitting spatial and velocity components (maser spots), it is highly probable that only the most compact component would remain on long space baselines. 
A raw SRT total power spectrum obtained in the first observing session is shown on Figure~2, where the $V_{\mathrm{LSR}}$ range corresponds to the 16~MHz upper side band. Note, that it shows a sinusoidal-like shape produced by an onboard digital seven-pole Butterworth filter. These specific "waves" with peak separations of about 1~MHz (corresponding to 13.5 km/s) do not significantly affect the cross power spectra, which do not have a response to the independent receiver noise contributions from the telescopes. 

Amplitude calibration of the data was made using SEFD (System Equivalent Flux Density)
measurements provided by ground observatories and SRT. 
%Note, that Tsys calibration measurements were made in \textit{on source} mode. 
But the difficulty here is that these a priori SEFD values are known with insufficient accuracy and, thus, scales for different telescopes differ from each other. In this case we have to calibrate the scales using the auto-correlation (AC) spectra of the maser emission that is commonly observed in all the telescopes and the SEFD value of the most well calibrated telescope as reference. 
However, the important thing is that the visibilities are calculated from the initial unnormalized autocorrelation spectra, i.e. they do not depend on the SEFDs.
This can be done because W49N is sufficiently strong and thus it is detectable with the SRT in its AC spectrum.

%Assuming that the Effelsberg 100-m (in the 1st and 2d sessions) and Green Bank 100-m (in the 3d session) provided high-quality Tsys measurements, we calibrated all the scales with respect to these telescopes. 
The SRT 10-m, Effelsberg 100-m and Green Bank 100-m total power spectra are shown on Figures~3-5. 
%The SRT total power spectrum is slightly cropped on both sides to remove the edge effects of the band. 
Note that the faintest velocity component which is clearly visible in the cross power SRT spectrum has a flux density of about 1000~Jy. The vector averaged cross power spectra corresponding to the whole frequency range of observations are also presented on Figures~3-5.

\section{Results and discussion}

It is well known that the H$_2$O maser emission in W49N shows strong and rapid variability with the bright high-velocity features sporadically arising in velocity range of a few hundred km/s \citep{Gwinn1994a}.
%The H$_2$O maser in W49N is known as a highly variable source displaying bright high-velocity features arising from time to time in velocity range of several hundred km/s.
Such features may have flux densities up to several thousand Jy and then disappear. The main part of the W49N H$_2$O maser spectrum around the systemic velocity at $\sim$0~km/s is always observable and had a peak flux density during our observations of 40~kJy. The low velocity components were in the observed band for both local oscillator settings. The main spectral masing features at $-6$~km/s had a total flux density of 45500, 11200 and 8200~Jy in the 1st, 2d and 3d observations, respectively. 
%Since \textit{RadioAstron} can observe in the bandwidth of 32~MHz, the central frequency in all observations of W49N was chosen so that the main part of the spectrum fell into the band. 

Further discussion on the number of maser components detected with space baselines is based on the visibility amplitude and phase profiles. We report only on features with a signal-to-noise ratio $\geq$7 on the SRT-ground baselines. The analysis of the distribution of all the W49N maser components based on the fringe-rate mapping method along with comparing it with previous results will be presented in a future publication.

Fringes on a space baseline of $\sim$3~ED between the SRT and 100-m Effelsberg telescope were found in the first observing session on 18 May 2014 for two spectral features at $-$6~km/s and at 6~km/s (see full range cross power spectrum on Figure~3). Lack of the intermediate length baselines does not allow us to explore the structure of the features so as to distinguish the flux density contributions from different structural components. 

In the second observing session on 27 April 2015, only the brightest component at the velocity $-$6~km/s was detected on the space baseline of 9.6 Earth diameters between the SRT and Effelsberg. The corresponding fringe spacing was 23~$\mu$as, which is the record fringe spacing achieved for galactic masers (see Figure~4, bottom panel). 
Along with the main part of the W49N spectrum a new short-lived maser feature at 64~km/s was observed in this session (see Figure~4, top panel). According to monitoring on the Effelsberg 100-m telescope \citep{Busaba2018}, this feature appeared a few weeks before the observation on \textit{RadioAstron} and disappeared a few months later. This short-lived component gave fringes only on ground baselines up to the longest baseline Effelsberg-Hartebeesthoek. The fringe on this baseline has a small amplitude of 4~Jy and is not readily seen on the scale of plotting. The possible feature at 64 km/s in the SRT-Effelsberg cross power spectrum does not show the normal constancy of phase across the profile, so we cannot confirm it to be a real detection.
%is a group of single channel noise spikes. 

During the third observation on 22 May 2015 a group of bright features around $-$63~km/s was observed (see Figure~5, top left panel), which produced fringes on space baseline of 8.6 Earth diameters between the SRT and Green Bank (GBT). Weak fringes were also found for a component at $-$72~km/s. No fringes on space baselines for other components were found in this session. 

The correlated fraction of the total flux density of the detected maser components are summarized in Table~2. It is very interesting that the most compact spectral components detected at space baselines in three experiments contain less than 1\% of the corresponding total flux density. The visibility amplitude vs the length of the projected baseline for the two compact features detected is plotted on Figure~6.

%Earlier the diffraction scattering sizes of maser spots for W49 which is set by the radiation propagation through the ISM have been reported in \citep{Gwinn1988}, see also \citep{Gwinn1994}. 
Gwinn et al.(1988a) analyzed the ground based visibility measurements of some features in W49N out to ground based baselines of 8000~km and concluded that the apparent angular sizes of $\sim$250~$\mu$as could be explained by Kolmogorov turbulence, presumably arising in the ISM (see also \citealt{Gwinn1988b}).
Our observations with \textit{RadioAstron} greatly extend the observing basis for the analysis of scattering effects. It is clearly seen from Figure 6 that the part of visibility curve corresponding to \textit{RadioAstron} measurements is sloping much more slowly than expected from the single Gaussian which is usually a good approximation for a source with the simple structure.
%It is clearly seen from Figure~6 that points corresponding to \textit{RadioAstron} measurements of the visibility are not a simple extrapolation of the short baseline data with a Gaussian function, for example. 
%We are investigating whether our measured low amplitude visibilities are due to reaching the refractive floor caused by interstellar scintillation when observing at such long baselines up to 9.6~ED (see, for example, \citealt{Johnson}). 
We are investigating whether our measured low amplitude visibilities are due to reaching the refractive noise level caused by interstellar scintillation when observing at such long baselines up to 9.6~ED. It was shown in \citep{Johnson} that at wavelength of 1.3~cm the rms refractive fluctuations (noise) in some cases could dominate measured visibilities at very long baselines of $\sim$120000~km and higher.
Without refractive scattering we probably would have seen no fringes on W49N with \textit{RadioAstron} because of the long propagation path through the ISM. 

An alternative explanation for the observed form of the visibility curve is that the maser spots probably have complex structure which results from the scattering by the ionized gas as proposed for W49N \citep{Gwinn1994b} or caused by the turbulence in the region where the maser is formed (see theoretical calculations by \citealt{Watson,Sobolev2008}). Influence of the turbulence on the ultra-fine structure of the maser images was described for the example of \textit{RadioAstron} observations of water maser emission in Cepheus~A by \citet{Sobolev2018}. Perhaps, both effects takes place, so this issue requires more investigation. 
%An alternative explanation for the observed visibilities is that they are caused by the turbulence in the region where the maser is formed as proposed for W49N by \citep{Gwinn1994b} and for Cepheus~A by \citep{Sobolev2018}. See also the theoretical calculations by \citep{Watson,Sobolev2008}.
%, similar to the case of Cepheus~A \citep{Sobolev2018}. 

\begin{table}
\scriptsize
\begin{center}
\centering
\caption{Correlated fraction of the total flux density of the most compact components of the W49N detected at space baselines in three observing sessions.}
\begin{tabular}{crcc}
\hline\hline
Observation & V$_\mathrm{LSR}$, & Fraction of the total & Baseline  \\
 & km/s & flux density & ED \\
\hline
18 May 2014   & $-$6 & 0.006 & 3.0 \\
              & $+$6 & 0.003 &  \\
\hline
27 Apr 2015   & $-$6 & 0.001 & 9.6 \\
\hline
22 May 2015   & $-$63 & 0.002 & 8.6 \\
              & $-$72 & 0.004 &  \\
\hline
\hline
\end{tabular}
\end{center}
\end{table}

\section{Conclusions}

\noindent In this paper we presented the results of observations of the water maser emission in W49N with the space interferometer \textit{RadioAstron} in the spectral mode in 2014-2015. A description of observing mode and general design of conducted sessions is provided. We also briefly described the spectral data processing procedure to show that it yields successful results not only for ground-based array, but also for space VLBI \textit{RadioAstron}.

The most compact maser spots in W49N were detected on space baselines up to 9.6 Earth diameters. The correlated flux density of the most compact structures is a fraction of a percent of the single dish flux density. These low values of correlated flux density are probably due to turbulence either in the maser itself or in the interstellar medium.

\section*{Acknowledgments}

The RadioAstron project is led by the Astro Space Center of the Lebedev Physical Institute of the Russian Academy of Sciences and the Lavochkin Scientific and Production Association under a contract with the State Space Corporation ROSCOSMOS, in collaboration with partner organizations in Russia and other countries.

This work is based on observations carried out using the 100-meter radio telescope of the MPIfR (Max-Planck-Institute for Radio Astronomy) at Effelsberg, 100-meter R.C.~Byrd Green Bank Telescope of the Green Bank Observatory, which is a facility of the National Science Foundation operated under cooperative agreement by Associated Universities, Inc., 26-meter radio telescope operated by Hartebeesthoek Radio Astronomy Observatory in Johannesburg, Republic of South Africa, 32-meter telescope in Medicina operated by INAF - Istituto di Radioastronomia and 32-meter radio telescope operated by Toru\'{n} Centre for Astronomy of Nicolaus Copernicus University in Toru\'{n} (Poland) and supported by the Polish Ministry of Science and Higher Education SpUB grant.

NNS acknowledges support from Russian Science Foundation grant 18-12-00193. AMS work was supported by the Ministry of Education and Science (the basic part of the State assignment, RK No. AAAA-A17-117030310283-7) and by the Act No. 211 of the Government of the Russian Federation, agreement 02.A03.21.0006.

%%%%%%%%%%%%%%%%%%%%%%%%%%%%%%%%%%%%%%%%%%%%%%%%%%%%%%%%%%%%%%%%%%%%%%%%%%%%%
%% Appendices
% \appendix

\clearpage

\begin{figure}
\label{uv}
\begin{minipage}[h]{0.95\linewidth}
\center{\includegraphics[width=0.8\linewidth]{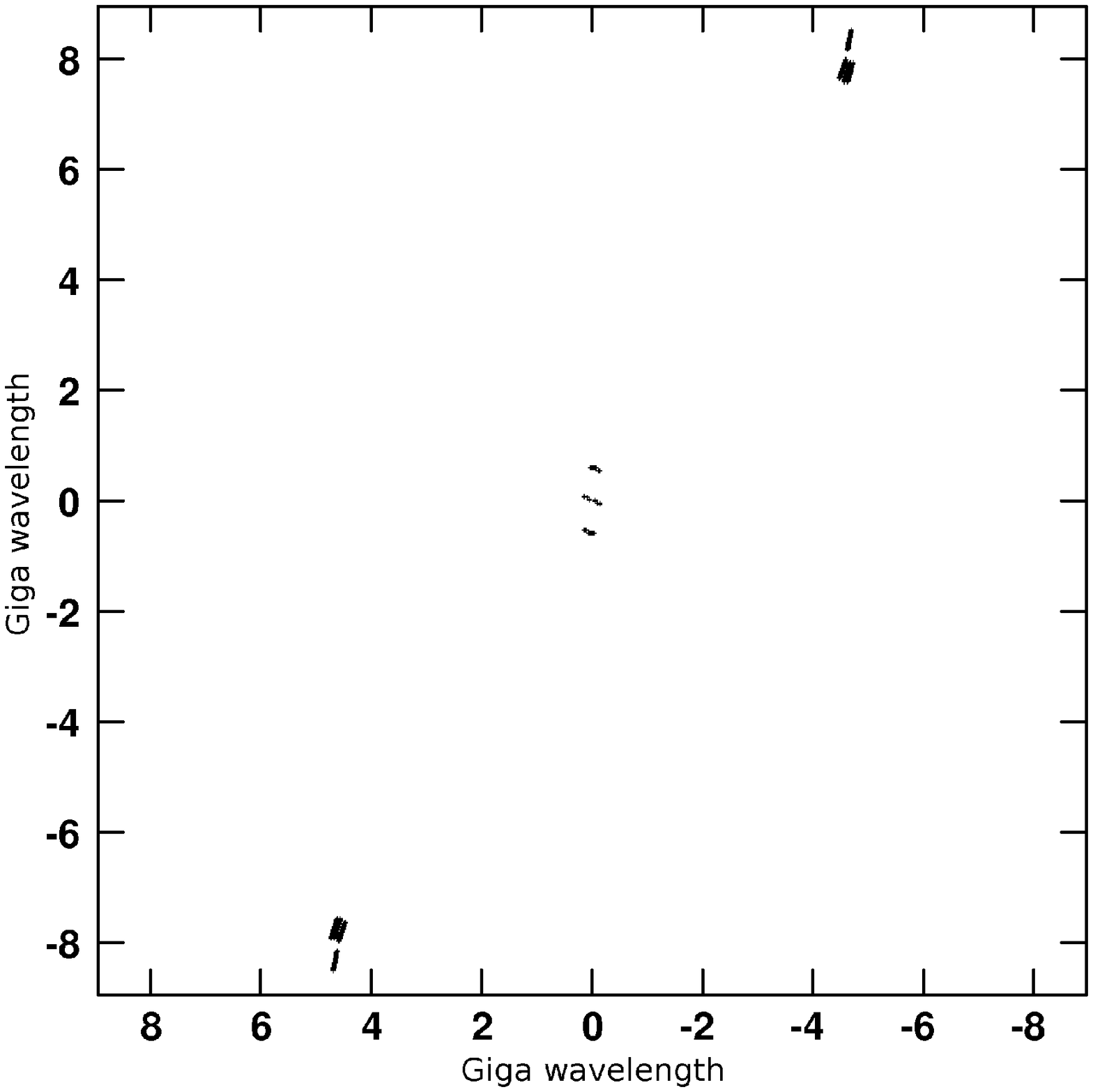}}
\end{minipage}
\vfill\vfill
\begin{minipage}[h]{0.49\linewidth}
\center{\includegraphics[width=1.0\linewidth]{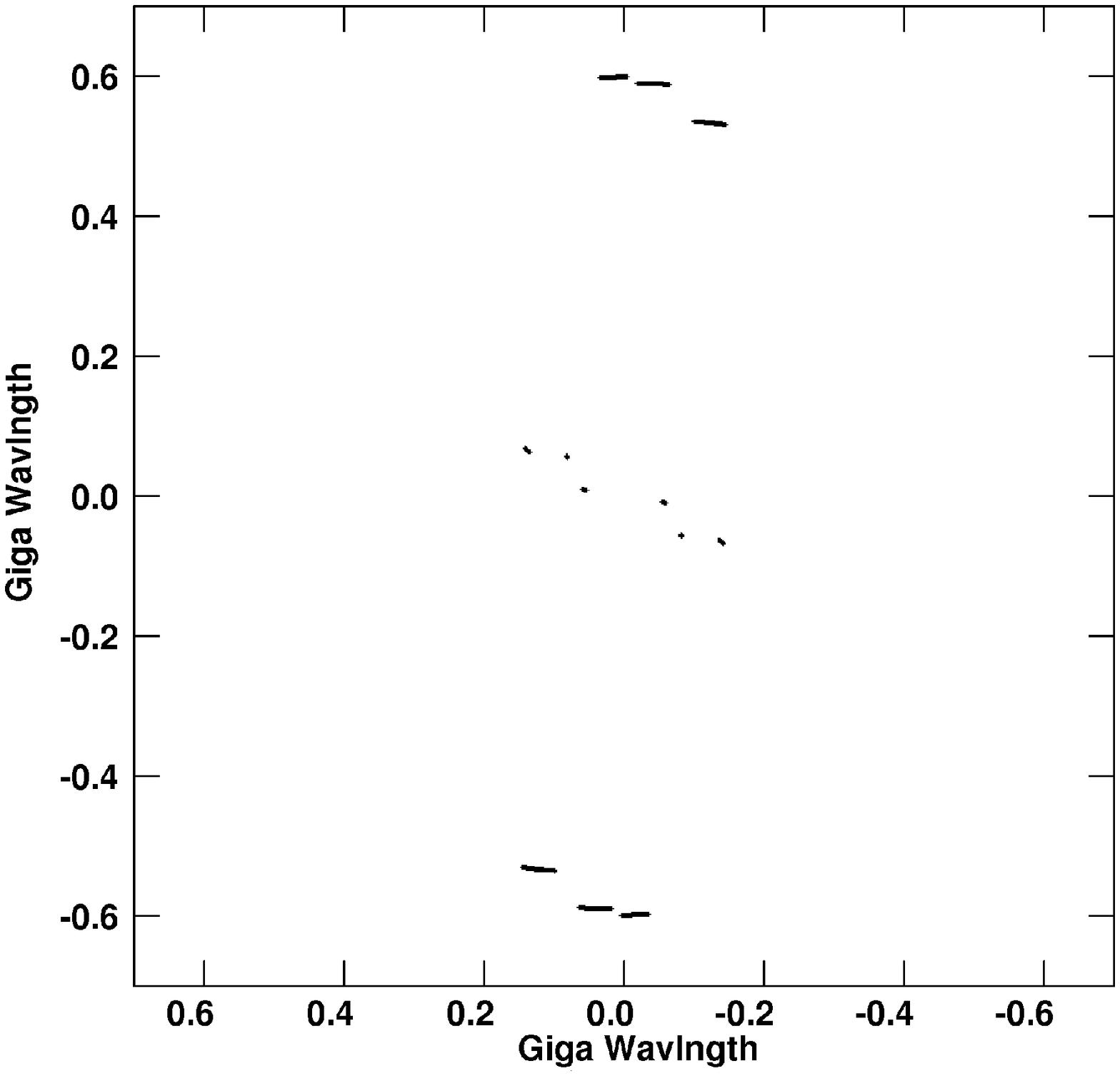}}
\end{minipage}
\hfill
\begin{minipage}[h]{0.49\linewidth}
\center{\includegraphics[width=1.0\linewidth]{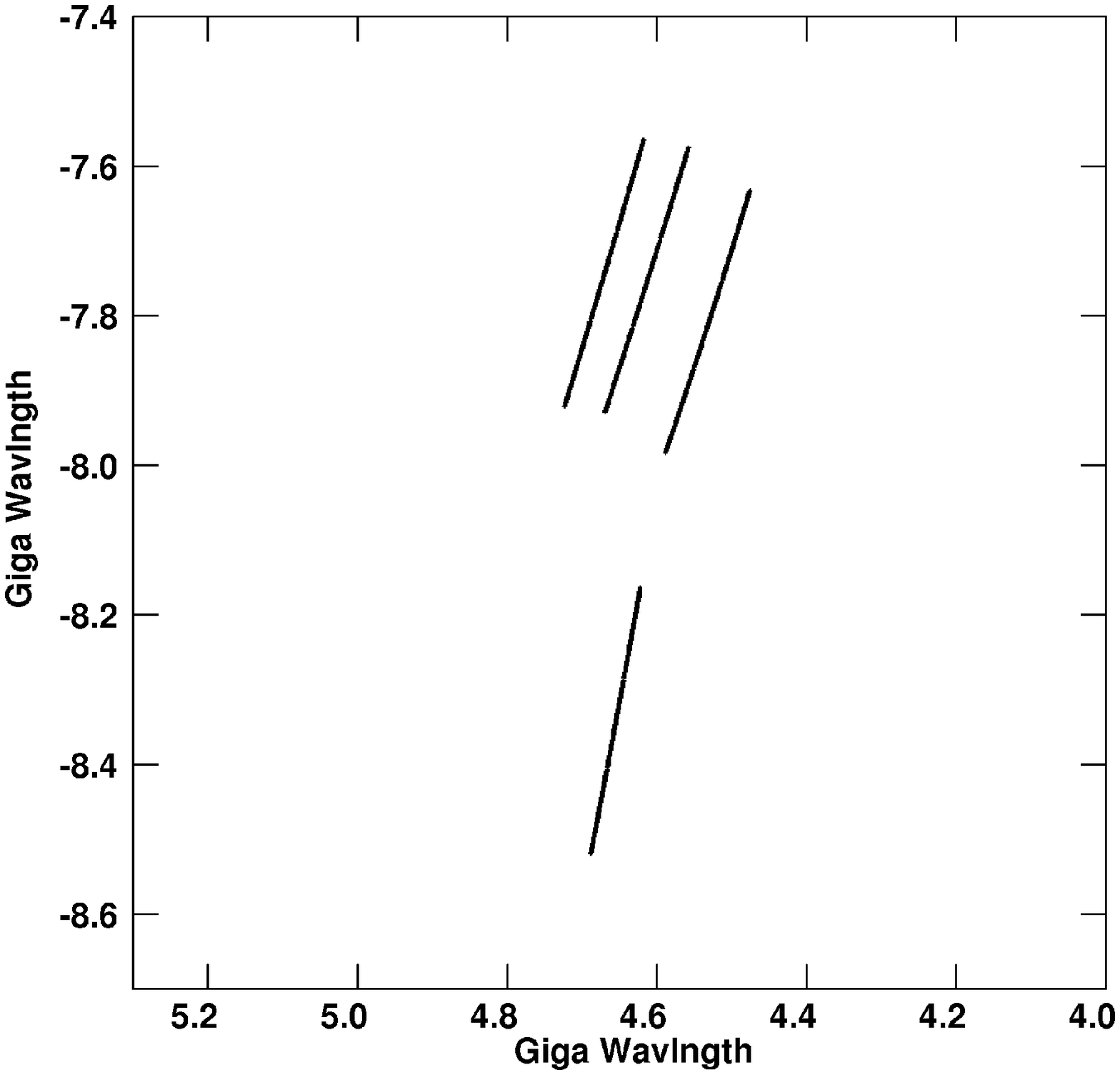}}
\end{minipage}
\centering{\caption{{\it Top panel}: uv-plane coverage for observing session on 27 April 2015, 01:00-02:00 UT. Participated telescopes: SRT 10-m, Effelsberg 100-m, Yebes 40-m, Toru\'{n} 32-m and Hartebeesthoek 26-m. {\it Bottom panel}: Expanded view of the uv-plane coverage for ground baselines ({\it left}) and one conjugate part of the space baselines (\textit{right}).}}
\end{figure}

\begin{figure}
\center{\includegraphics[width=1.0\linewidth]{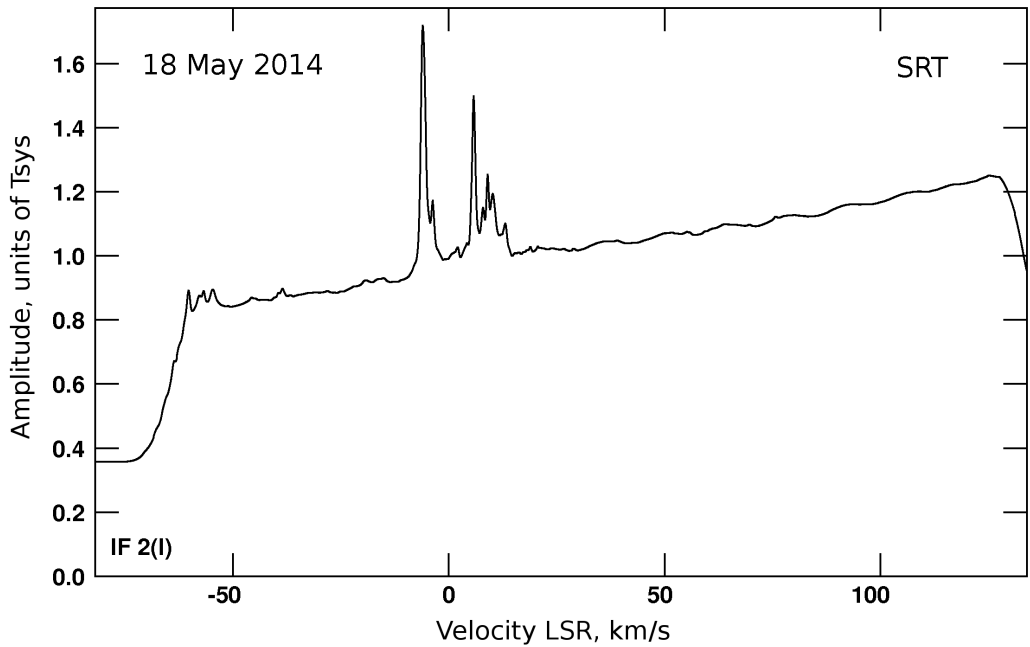}}
\vfill
\centering{\caption{Raw SRT total spectrum (Stokes I) of W49N in units of system temperature obtained in the \textit{RadioAstron} observation on 18 May 2014. V$_\mathrm{LSR}$ range corresponds to the 16 MHz upper side band (frequency range 22.228$-$22.244~GHz). This figure shows that the peak antenna temperature on the maser was almost equal to the system temperature of about 130~K. Note that the base band filter has a small ripple of period about 14 km/s.}}
\end{figure}

\begin{figure}
\centering{
\begin{minipage}[h]{0.6\linewidth}
\includegraphics[width=1.0\linewidth]{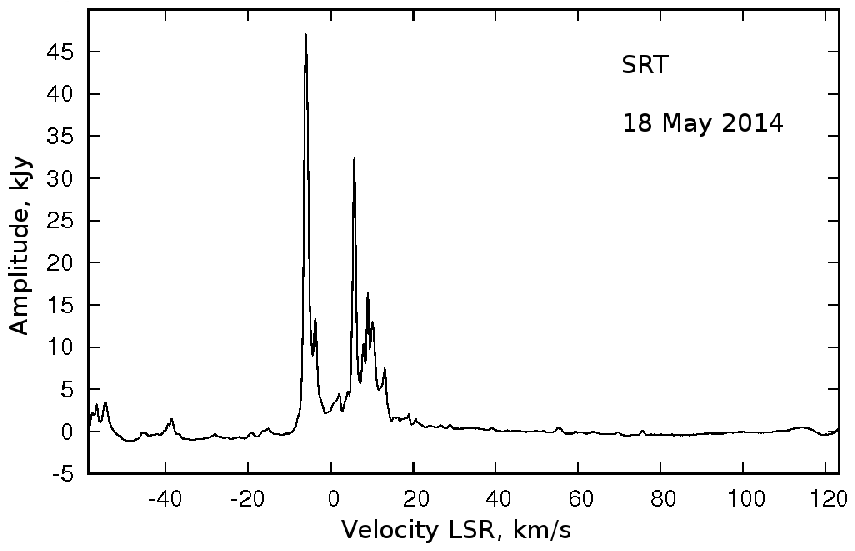}
\end{minipage}
\vfill
\begin{minipage}[h]{0.6\linewidth}
\center{\includegraphics[width=1.0\linewidth]{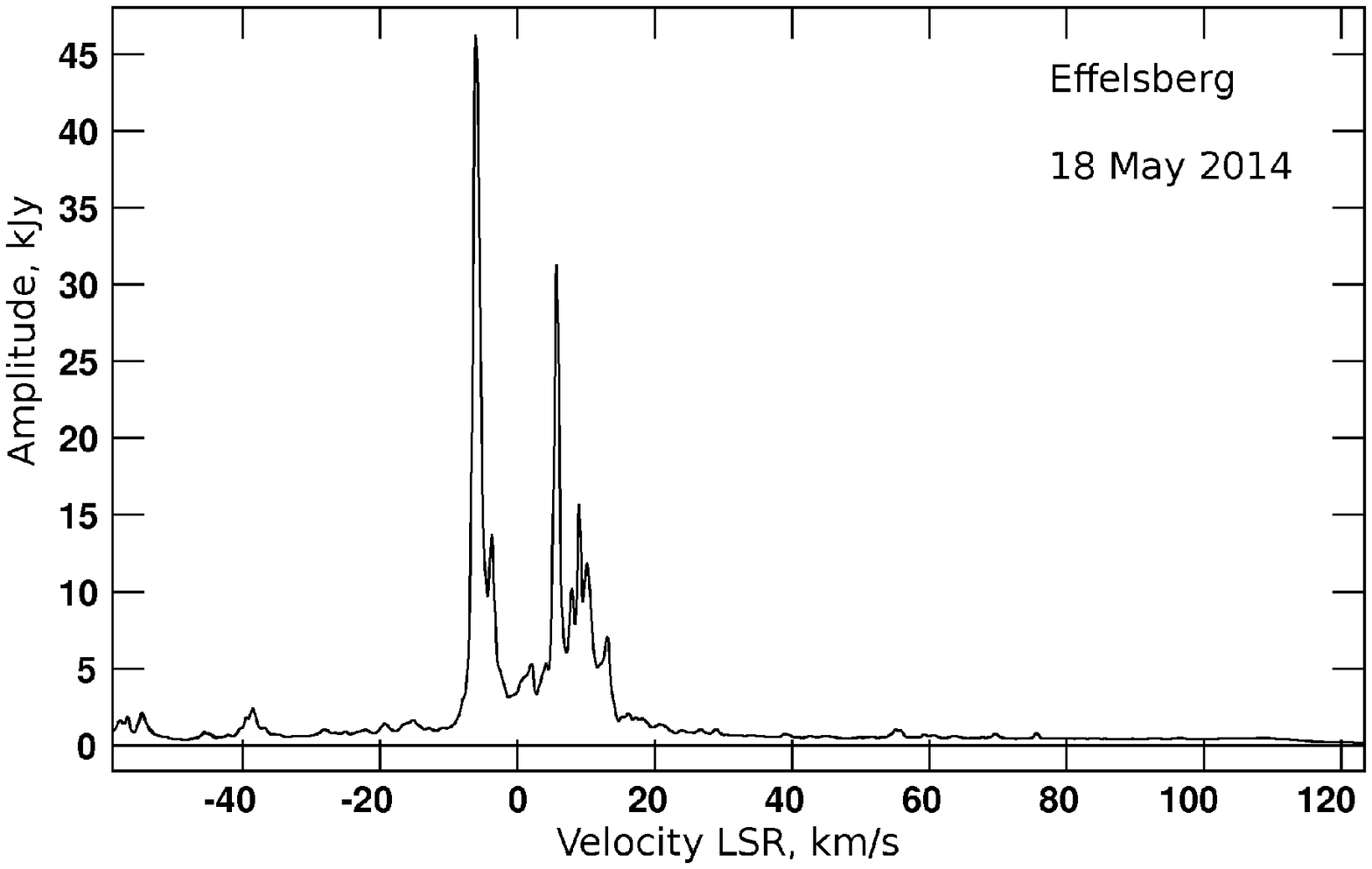}}
\end{minipage}
\vfill
\begin{minipage}[h]{0.6\linewidth}
\center{\includegraphics[width=1.0\linewidth]{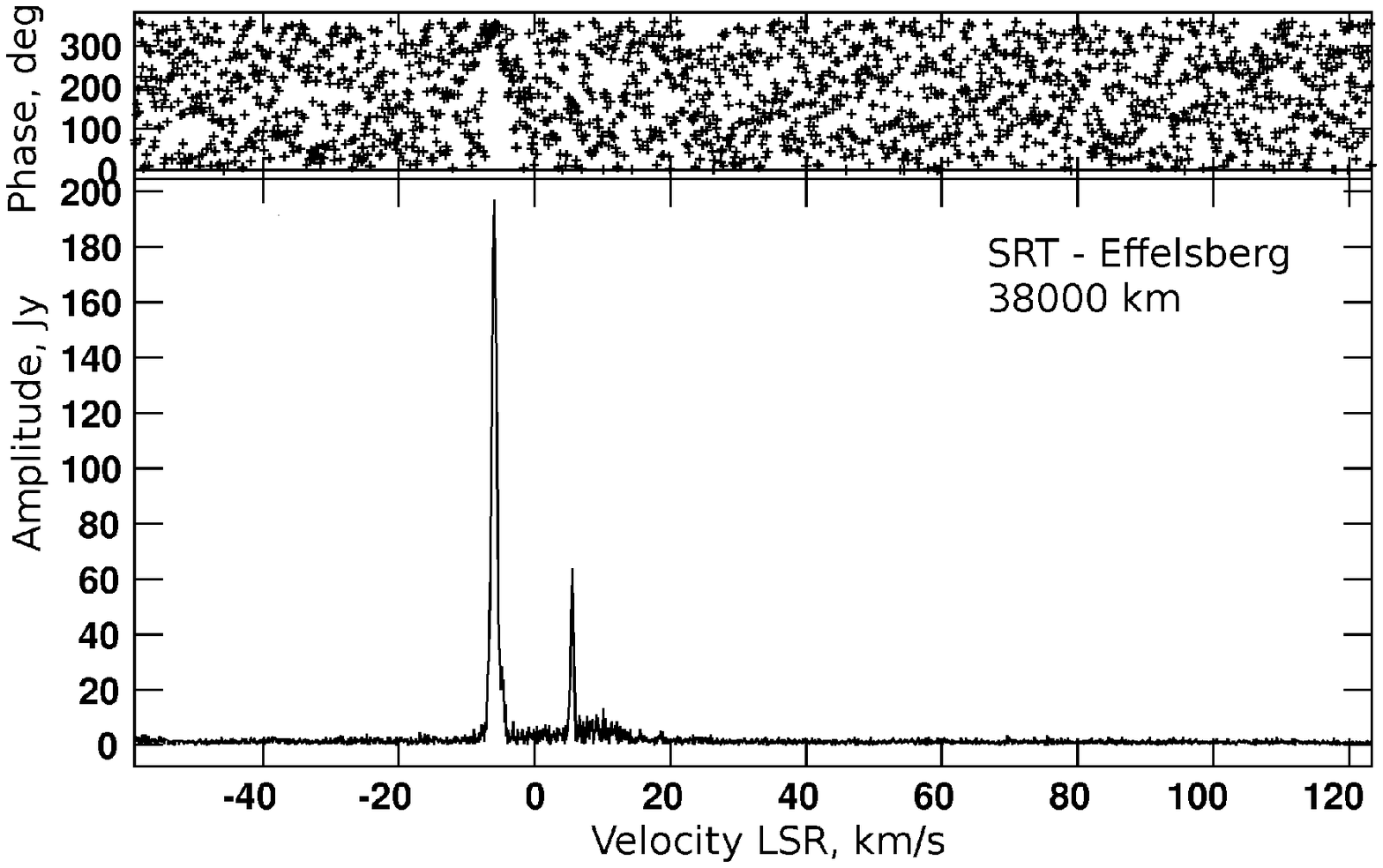}}
\end{minipage}}
\vfill
\centering{\caption{Autocorrelation spectra of W49N obtained on the 10-m SRT with a baseline removed (\textit{top panel}) and 100-m Effelsberg (\textit{middle panel}) in the \textit{RadioAstron} observation on 18 May 2014. V$_\mathrm{LSR}$ range corresponds to the 16 MHz upper side band (frequency range 22.228$-$22.244~MHz. \textit{Bottom panel}: cross power spectrum obtained at SRT-Effelsberg baseline. Telescopes and corresponding length of projected baseline are indicated on the figures. Note that the noise level in the part of the cross power spectrum (bottom) around 0 and 10~km/s is obviously higher 
than in the part of spectrum, where the source is weak ($<-20$~km/s and $>30$~km/s). This is due to the enhanced system temperature due to the high source flux density levels as can be inferred from the middle panel.}}
\end{figure}

\begin{figure}
%\label{spectrarags11ar}
\centering{
\begin{minipage}[h]{0.8\linewidth}
\center{\includegraphics[width=1.0\linewidth]{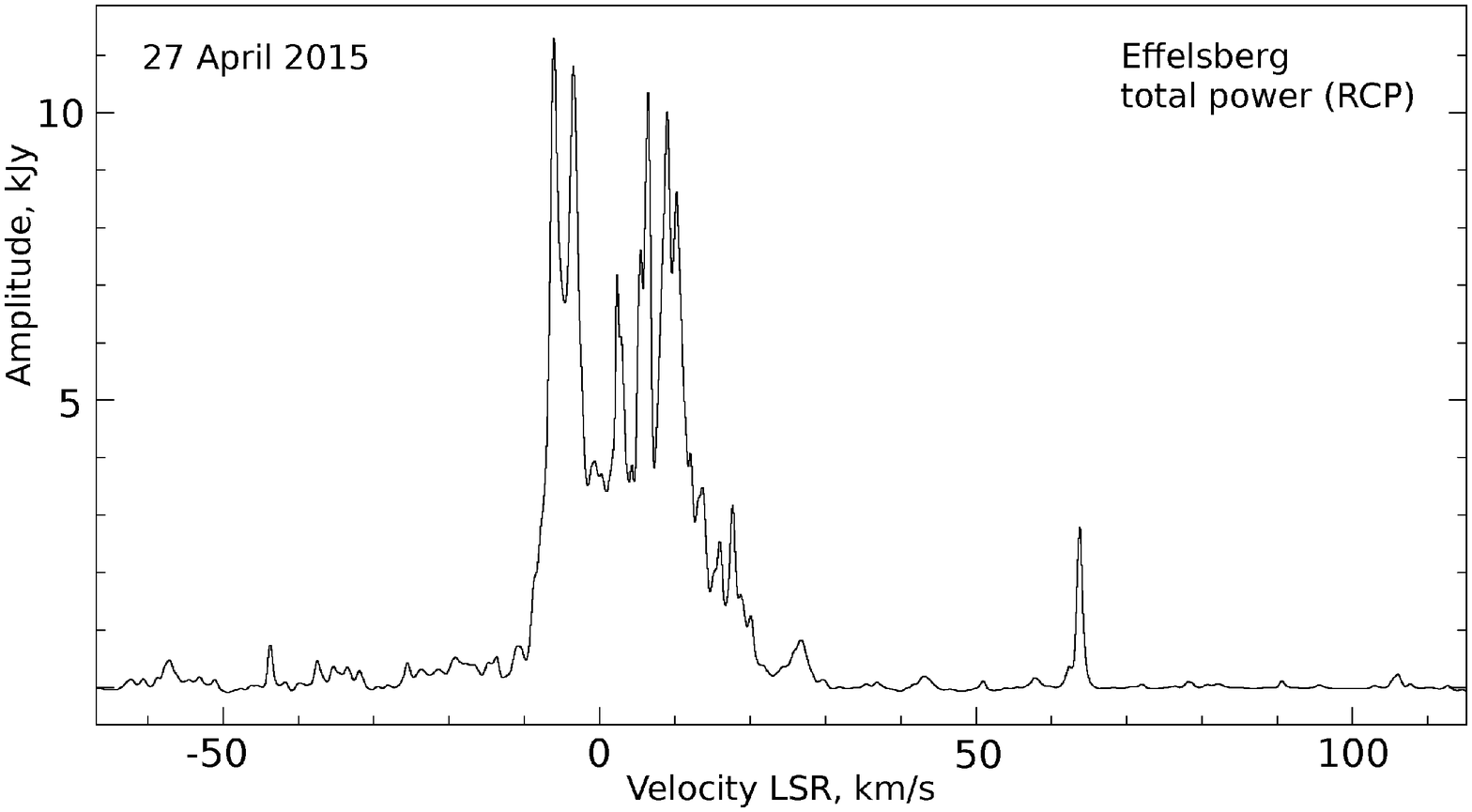}}
\end{minipage}
\vfill
\begin{minipage}[h]{0.49\linewidth}
\center{\includegraphics[width=1.0\linewidth]{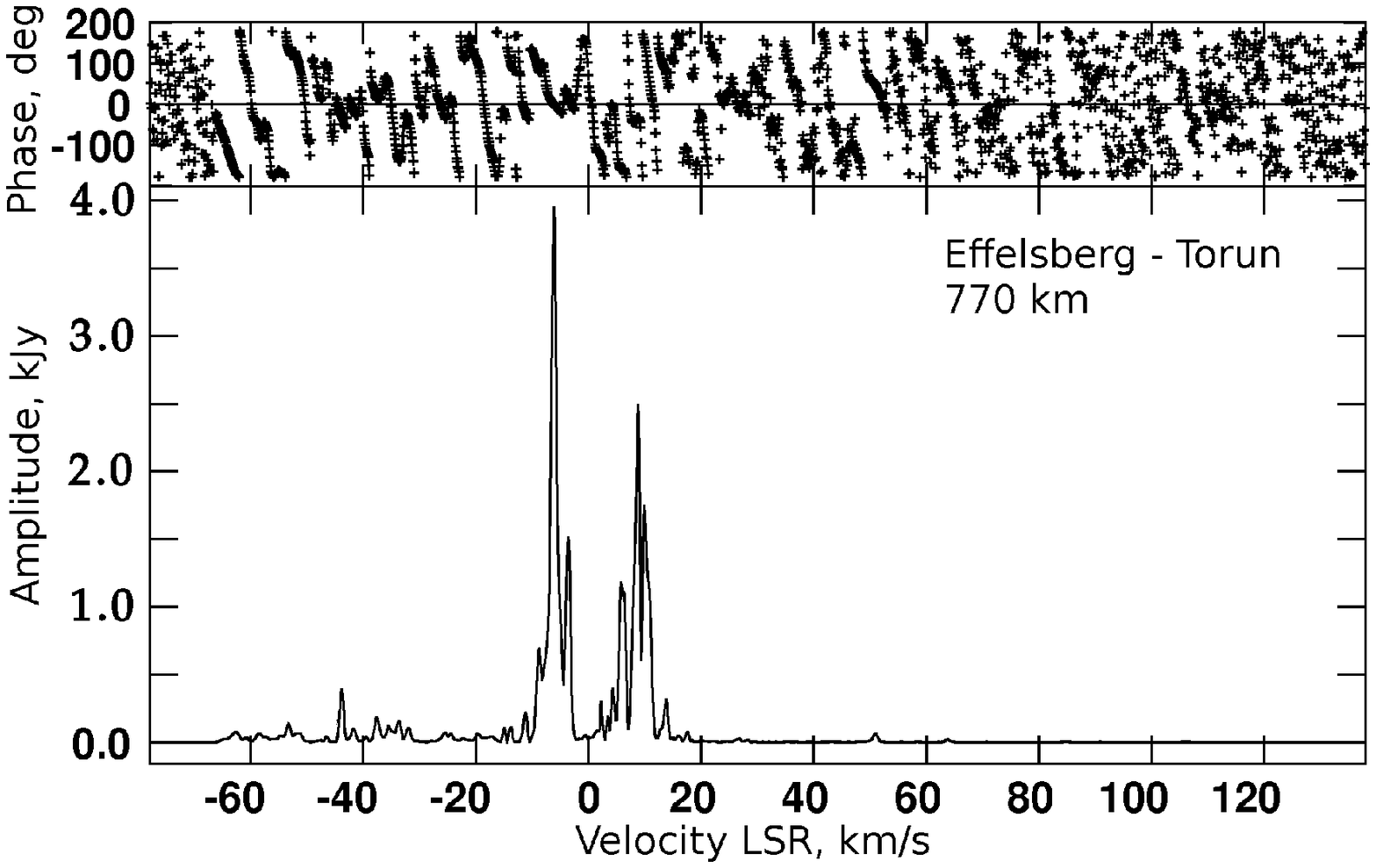}}
\end{minipage}
\hfill
\begin{minipage}[h]{0.49\linewidth}
\center{\includegraphics[width=1.0\linewidth]{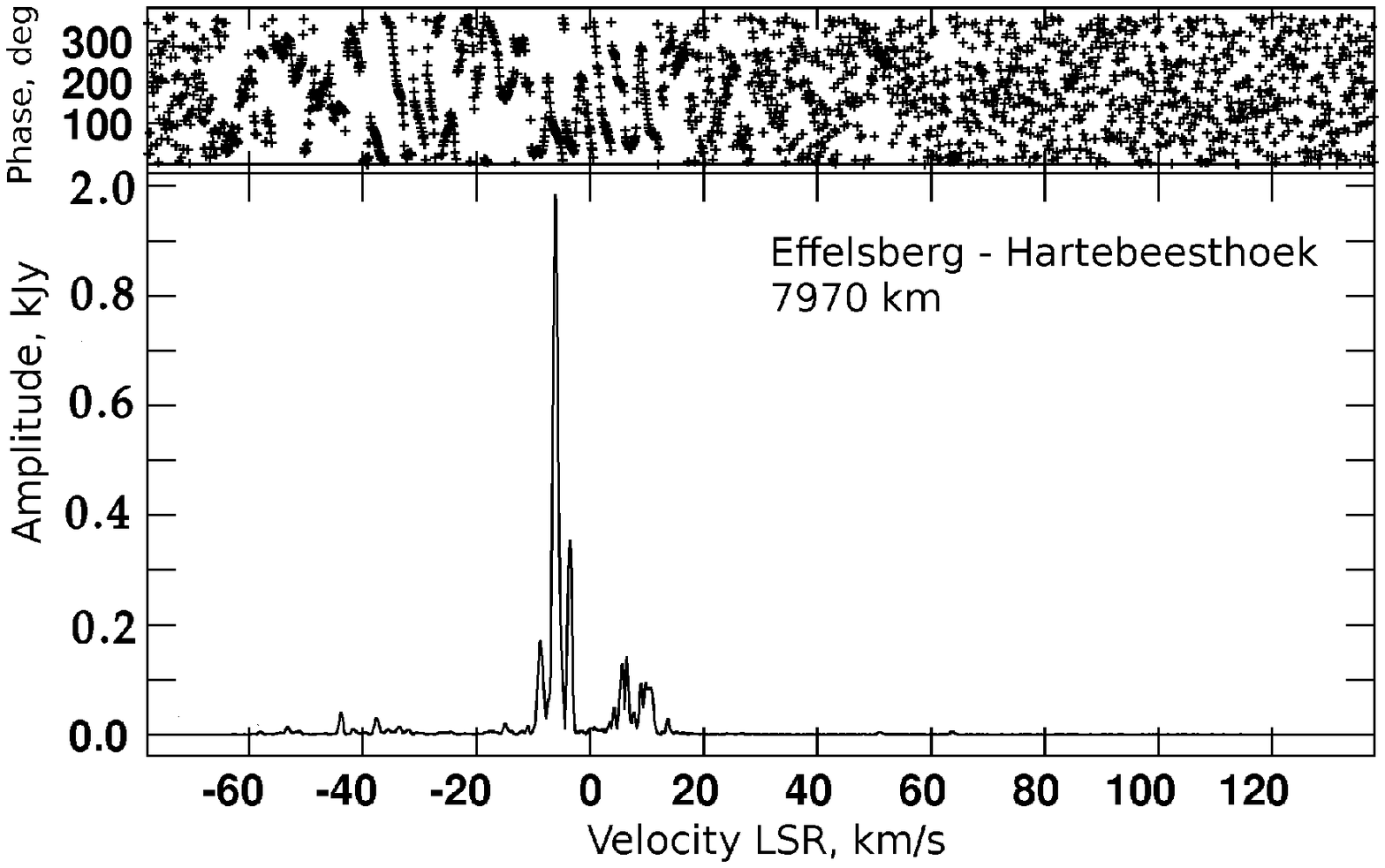}}
\end{minipage}
\vfill
\begin{minipage}[h]{0.49\linewidth}
\center{\includegraphics[width=1.0\linewidth]{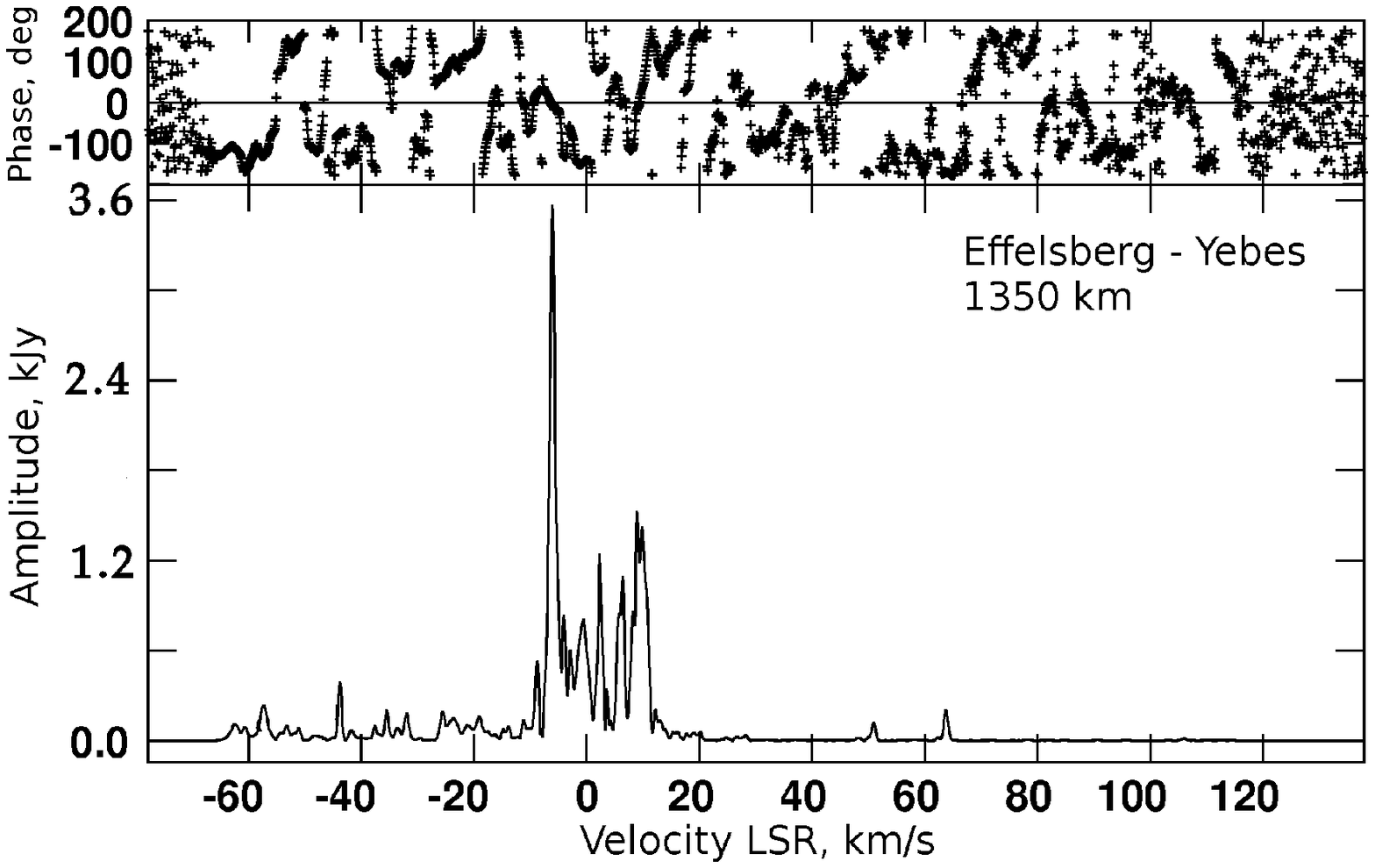}}
\end{minipage}
\hfill
\begin{minipage}[h]{0.49\linewidth}
\center{\includegraphics[width=1.0\linewidth]{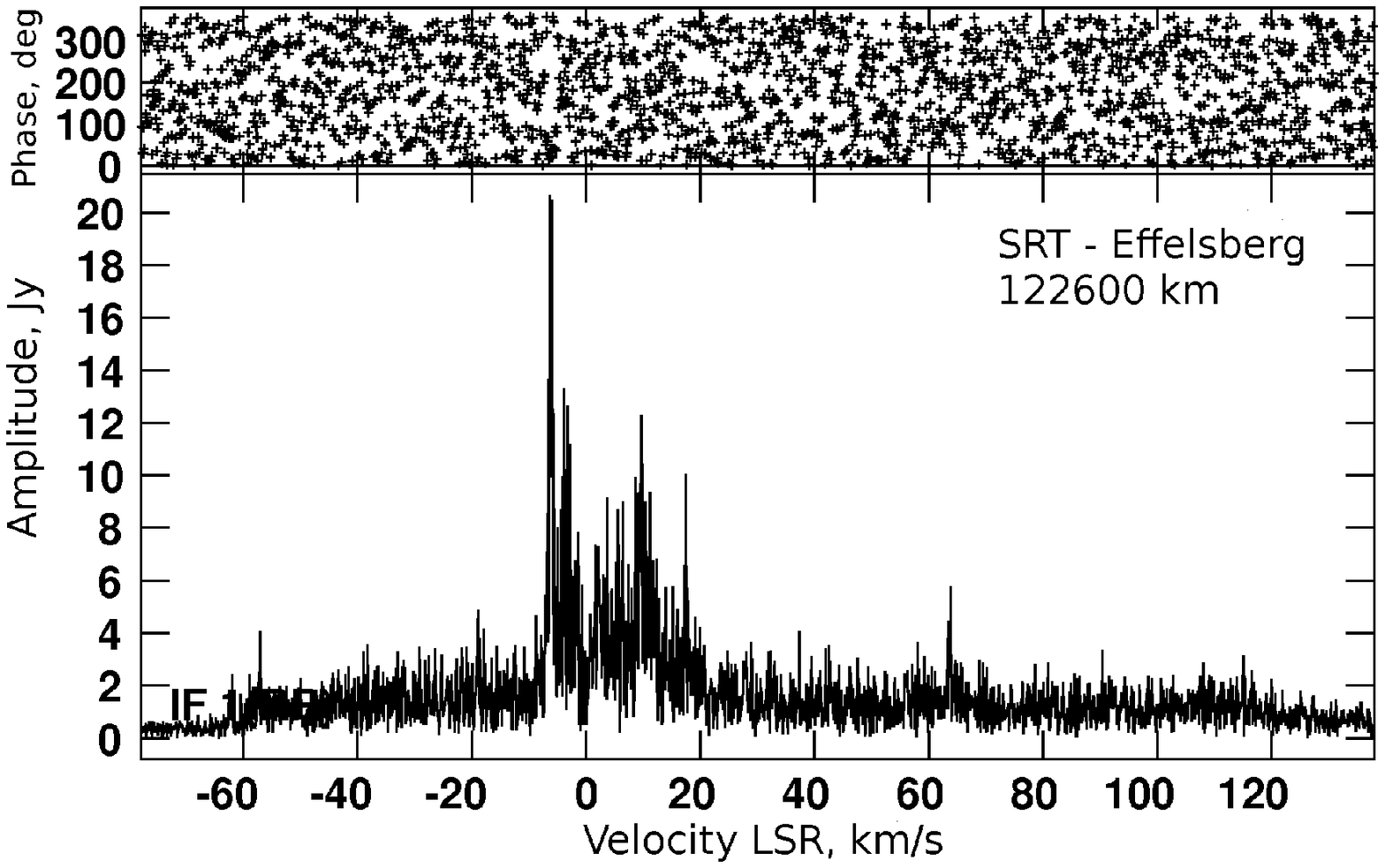}}
\end{minipage}
\vfill}
\centering{\caption{Total power spectrum (\textit{top panel}) and cross power spectra (\textit{middle} and \textit{bottom panels}) of W49N obtained in the \textit{RadioAstron} observation on 27 April 2015. V$_\mathrm{LSR}$ range corresponds to the 16~MHz upper side band (frequency range 22.228$-$22.244~MHz). 
Corresponding length of projected baselines and telescopes are indicated on the figures.}}
\end{figure}

\begin{figure}
%\label{spectrarags11ay}
\begin{minipage}[h]{0.5\linewidth}
\center{\includegraphics[width=1.0\linewidth]{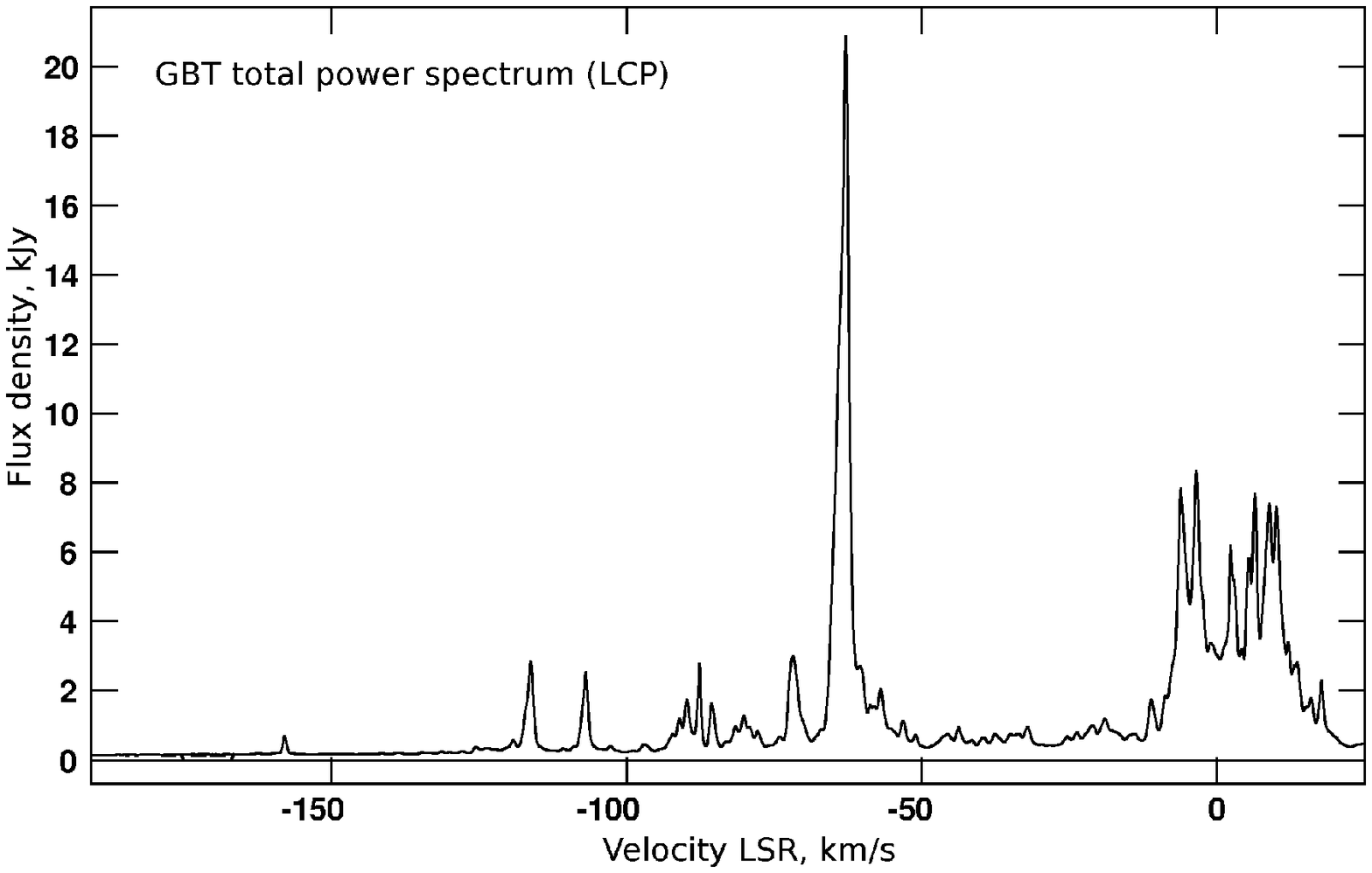}}
\end{minipage}
\hfill
\begin{minipage}[h]{0.5\linewidth}
\center{\includegraphics[width=1.0\linewidth]{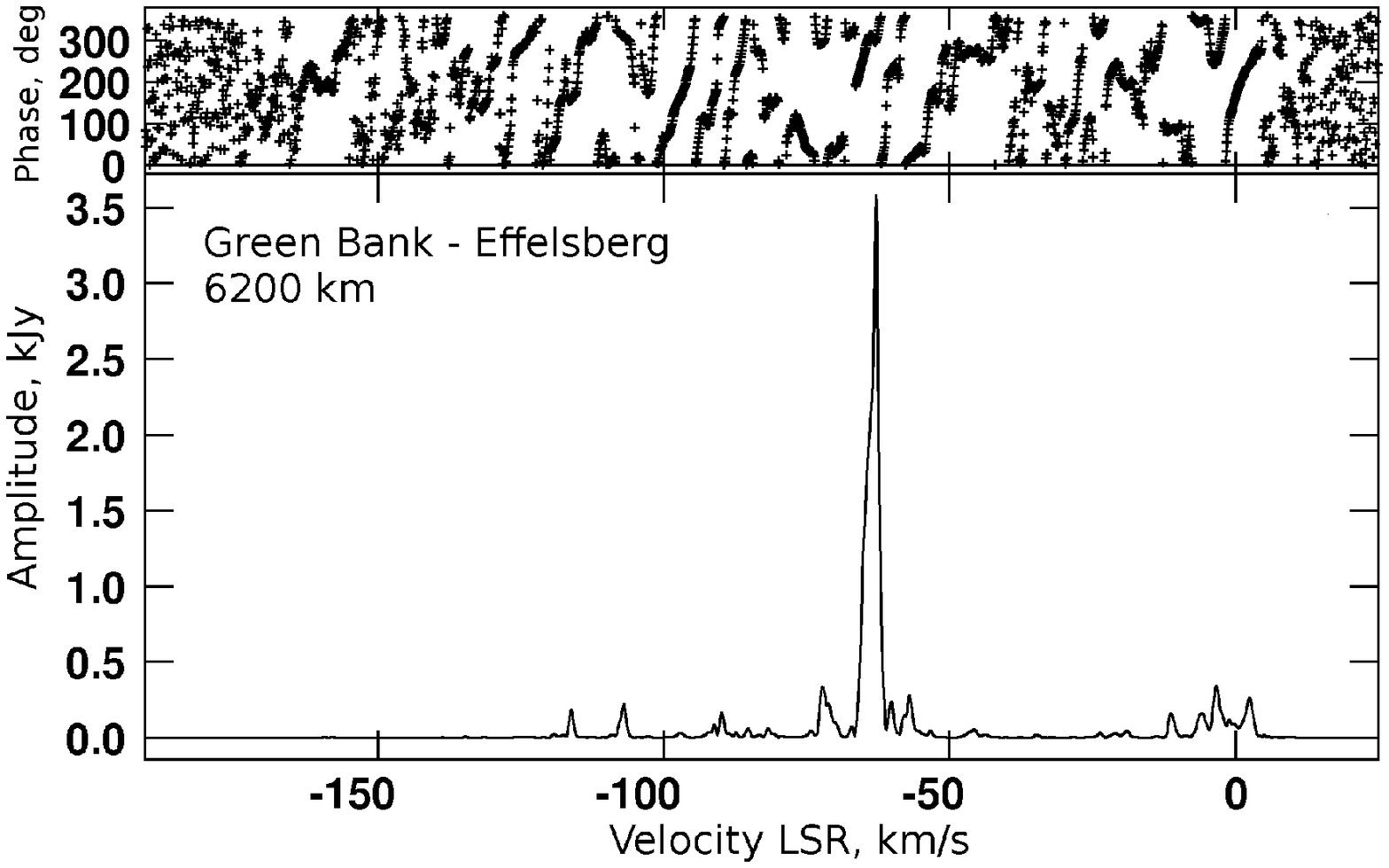}}
\end{minipage}
\vfill
\begin{minipage}[h]{0.5\linewidth}
\center{\includegraphics[width=1.0\linewidth]{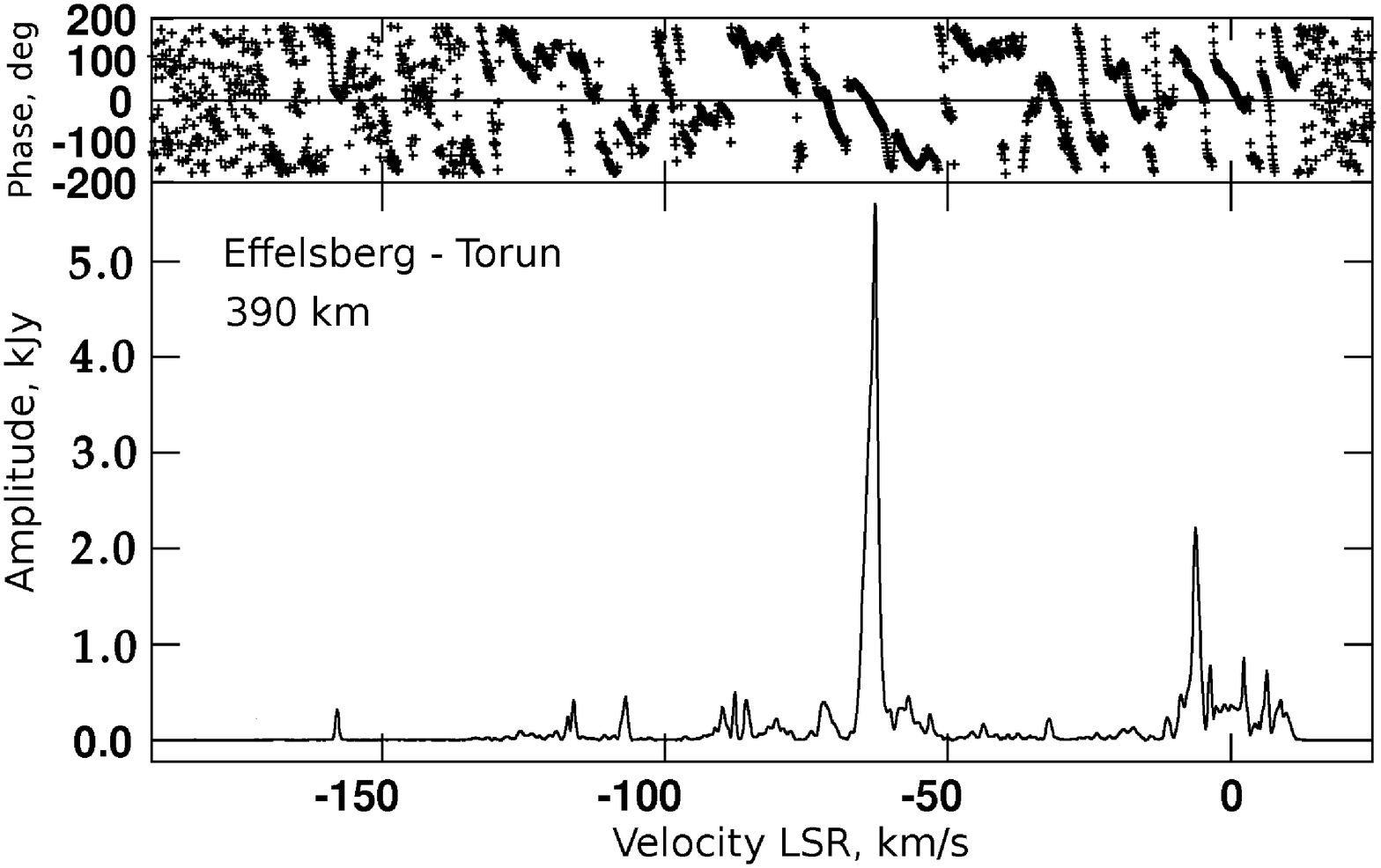}}
\end{minipage}
\hfill
\begin{minipage}[h]{0.5\linewidth}
\center{\includegraphics[width=1.0\linewidth]{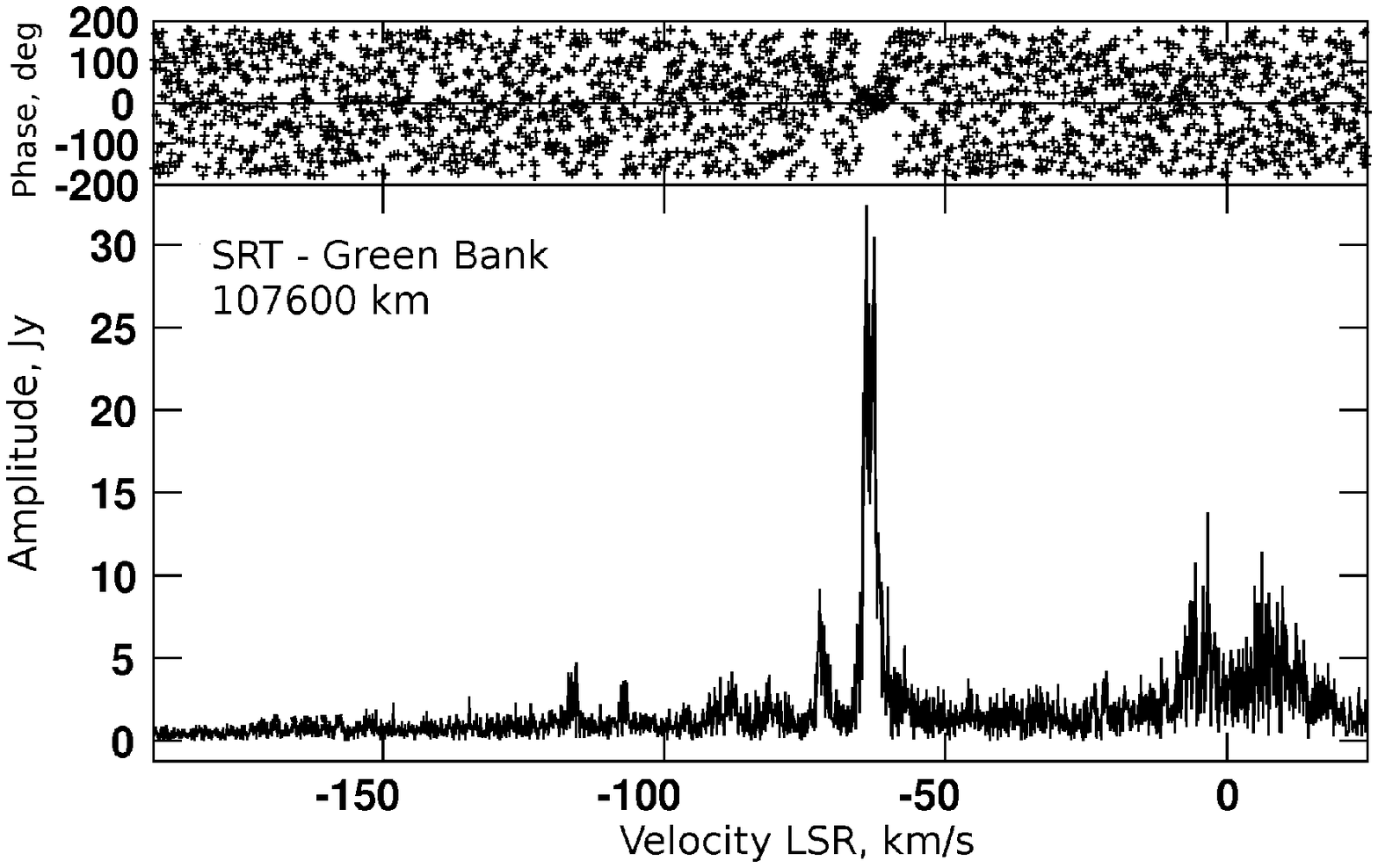}}
\end{minipage}
\vfill
\centering{\caption{Autocorrelation spectrum (\textit{top left panel}) and cross-correlation spectra (\textit{top right} and \textit{bottom panels}) of W49N obtained in the \textit{RadioAstron} observation on 22 May 2015. V$_\mathrm{LSR}$ range corresponds to the 16~MHz upper side band (frequency range 22.236$-$22.252~MHz). Corresponding length of projected baselines and telescopes are indicated on the figures.}}
\end{figure}

\begin{figure}
\center{\includegraphics[width=1.0\linewidth]{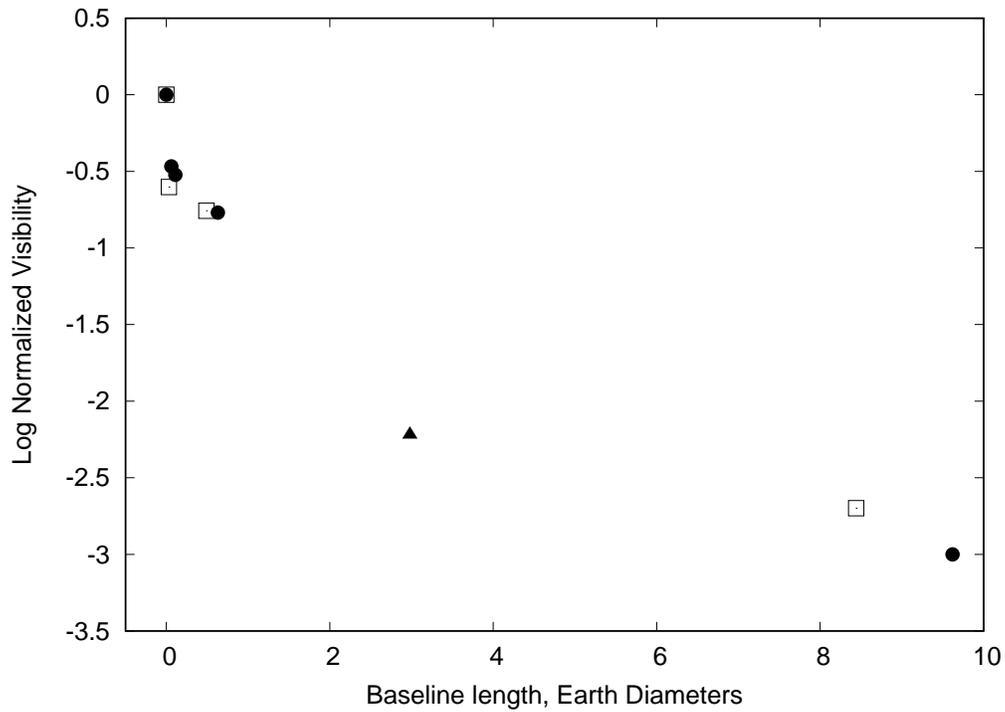}}
\centering{\caption{Normalized visibility for the two compact components observed in the two corresponding \textit{RadioAstron} experiments. Black points indicate visibility measurements for the most compact component at V$_\mathrm{LSR}=-6$~km/s detected in the 1st observing session (triangle) and in the 2d session (circles). Empty squares indicate visibilities for the feature at V$_\mathrm{LSR}=-63$~km/s detected in the third session. We adopt an accuracy of 10\% based on systematic effects, since the random effects are small because of our high signal-to-noise ratio.}}
\end{figure}

\end{document}